\documentclass{article}
\usepackage[utf8]{inputenc}
\usepackage[T1]{fontenc}
\usepackage{geometry}
\usepackage{float}
\usepackage{color}
\usepackage{array}
\usepackage{multirow}
\usepackage{amstext}
\usepackage{graphicx}
\usepackage{nicefrac}
\usepackage{inputenc}
\usepackage[spanish]{babel}
\usepackage{authblk}
\usepackage{ulem}

\date{}

\begin{document}

\title{Local and global stability analysis of a Curzon--Ahlborn model applied to power plants working at maximum $k$--efficient power}

\author[$1,*$]{G. Valencia--Ortega}
\author[$1$]{S. Levario--Medina}
\author[$2$]{M. A. Barranco--Jim\'enez}
\affil[$1$]{Departamento de F\'{i}sica, Escuela Superior de F\'{i}sica y Matem\'{a}ticas, Instituto Polit\'{e}cnico Nacional, U. P. Zacatenco, Edif. 9, 2o Piso, Ciudad de M\'{e}xico, 07738, M\'{e}xico.}
\affil[$2$]{Escuela Superior de C\'{o}mputo del Instituto Polit\'{e}cnico Nacional, Av. Miguel Bernard, Esq. Av. Miguel Oth\'{o}n de Mendizabal, Colonia Lindavista, Ciudad de M\'{e}xico 07738, M\'{e}xico.

$^{1,*}$gvalencia@esfm.ipn.mx
$^1$levario@esfm.ipn.mx
$^2$mbarrancoj@ipn.mx}

\maketitle

\begin{abstract}
The analysis of the effect of noisy perturbations on real heat engines, working on any steady--state regime has been a topic of interest within the context of Finite-Time Thermodynamics (FTT). The study of their local stability has been proposed through the so--called performance regimes: maximum power output, maximum ecological function, among others. Recently, the global stability analysis of an endoreversible heat engine was also studied taking into account the same performance regimes. We present a study of local and global stability analysis of power plant models (the Curzon--Ahlborn model) operating on a generalized efficient power regime called maximum k-efficient power. We apply the Lyapunov stability theory to construct the Lyapunov functions to prove the asymptotically stable behavior of the steady-state of intermediate temperatures in the Curzon--Ahlborn model. We consider the effect of a linear heat transfer law on the phase portrait description of real power plants, as well as the role of the $k$ parameter in the evolution of perturbations to heat flow. In general, restructured operation conditions show better stability in external perturbations. 
\end{abstract}

\section{Introduction}
The study of stability and dynamics robustness of heat engines models continues to be a topic of interest to establish optimal operating conditions, which preserve the steady--state regimes when the effect of external perturbations is considered \cite{GonzAyaletal19,GonzSantill18,klu,ChenThermal2018}. Since the paper published by Curzon and Ahlborn \cite{CAMod} and the discipline called finite-time thermodynamics (FTT) emerged \cite{DeVosLibro,Hoffmann97,Durmayaz04,SalamonLibro,Wu99}, several Curzon--Ahborn (CA) type thermal engine models have been studied through different operation modes that correspond to different objective functions such as maximum power \cite{CAMod,DeVosLibro}, efficient power \cite{Yilmaz,JEI2009}, ecological function \cite{Angulo91} and omega function \cite{Calvo01}, among others. Most studies within the context of FTT have focused on studying the steady--state energy properties of these objective functions. In addition, optimal values need to be associated to both design and construction parameters involved in thermal engine models to fulfill operation modes. Recently, Levario--Medina et al \cite{Levario1} proposed a new operating regime called $k$--efficient power. By using the extremal properties of $k$--efficient power, the authors found the best performance conditions in terms of the design and construction parameters for each energy converter (power plant); so each one can be operated in an energetic zone characterized by high power output and high efficiency.

Since Santill\'an et al \cite{Santillan2001} studied the local stability of an endoreversible CA heat engine operating at maximum power output regime, several authors have analyzed the role of external perturbations on the control parameters. For instance, the effect of the heat transfer laws and the thermal conductances on the local stability of the same endoreversible heat engine was investigated in \cite{Guzmanetal}. The local stability of a non--endoreversible CA model, taking into account the engine’s time delays operating at maximum power regime was also analyzed \cite{Paezetal}. In addition, the local stability of a heat engine model by considering some economic aspects related to the total cost of the heat engine operation as well as different performance regimes \cite{Barrancoetal1,Barrancoetal2} was studied. Other developments about local stability have been carried out even considering non linear heat transfer laws in the CA model \cite{klu,Wuetal,Chenetal,Wu}. Furthermore, other studies on local stability of models for heat pumps and refrigerators have been analyzed \cite{Huang1,Huang2,PAN2017}. More recently, Keune et al \cite{Keune2020} studied the stability of an absorption refrigeration powered by a wood boiler. All of the aforementioned studies have been focused on determining the relaxation times of the decaying rate to the stationary fixed points, assuming local perturbations around these values. On the other hand, Reyes--Ram\'irez et al \cite{I1,I2} investigated the global stability of a CA heat engine operating at different performance regimes by means of the Lyapunov method. They found the Lyapunov function as a way to prove the asymptotic stability behavior of the intermediate temperatures around the steady state. In this work, following the procedure reported in \cite{I1,I2}, we study the local and global stability of the $k$--efficient power regime \cite{Levario1}. The paper is organized as follows: In Section 2, we present the main steady state characteristics of an endoreversible engine model (CA model) at maximum $k$--efficient power regime. In Section 3, we describe the local stability analysis method for a set of power plants working at maximum $k$--efficient power. In Section 4, we explain the global stability analysis method based on Lyapunov's theory to construct the Lyapunov functions for the same set of power plants. Finally, in Section 5, we present our conclusions.

\section{Steady states of a CA heat engine (endoreversible model)}

\begin{figure}
    \centering
    \includegraphics[scale=0.5]{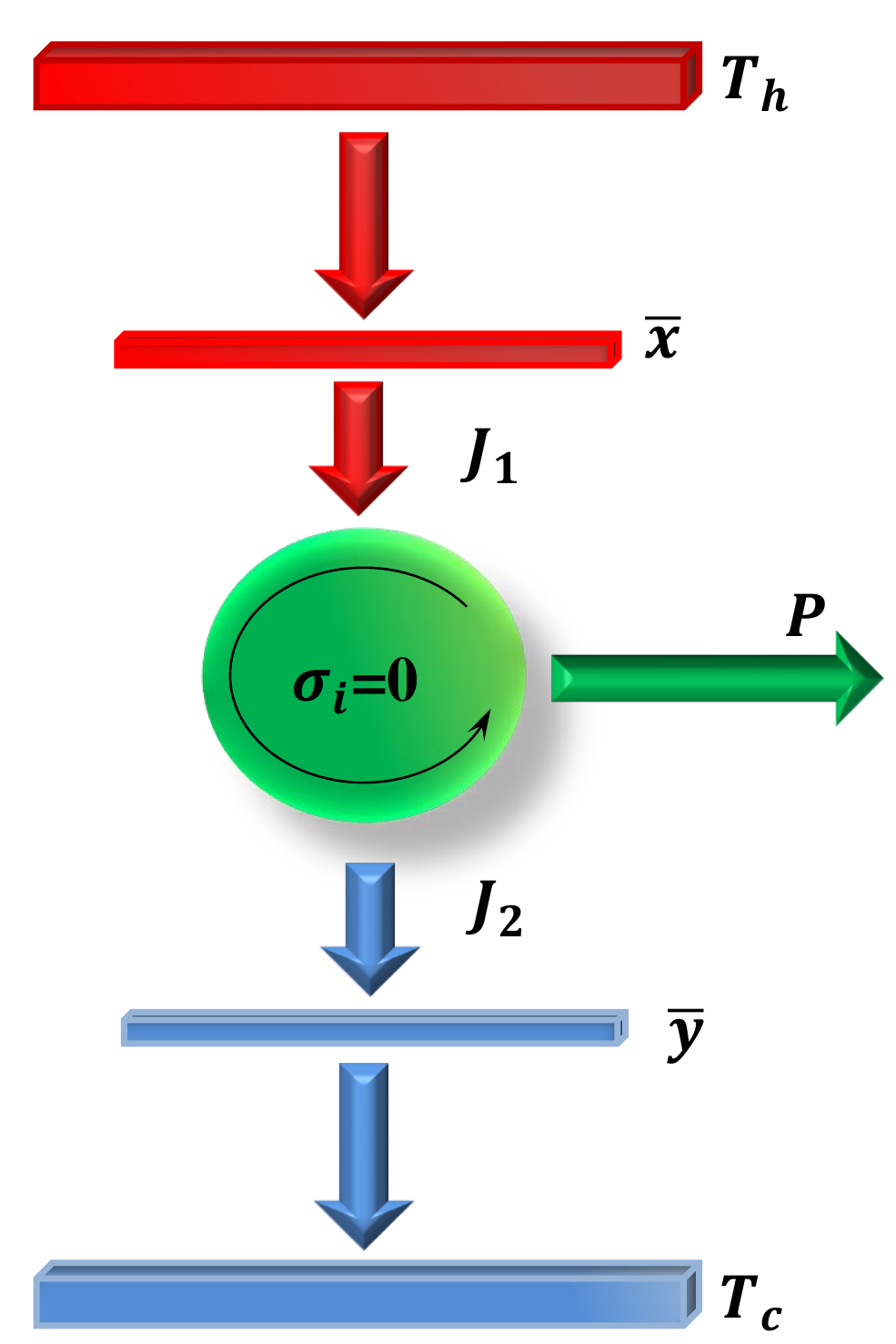}
    \caption{Scheme of an endoreversible Curzon-Ahlborn heat engine.}
    \label{fig:CAeng40}
\end{figure}

The typical model of a CA heat engine (see Fig. \ref{fig:CAeng40}) represents a working fluid operating in cycles between external reservoirs $T_h$ and $T_c$, with $T_h > T_c$. The energy dissipated between the working fluid and the reservoirs is represented by a heat transfer law (heat exchangers). $T_{hw}=\bar{x}$ and $T_{cw}=\bar{y}$ are known as internal working temperatures, and they define heat fluxes $\dot{Q}_{h}=J_h$ (from the internal heat deposit at temperature $\bar{x}$ to the system) and $\dot{Q}_{c}=J_c$ (from the system to the internal heat deposit at temperature $\bar{y}$). Hence, we use $\bar{x}$, $\bar{y}$, $\bar{J_h}$ and $\bar{J_c}$ to specify temperatures and heat fluxes in a steady state \cite{Santillan2001,Guzmanetal,Paezetal,Barrancoetal1,Barrancoetal2,I1,I2}. As the endoreversibility hypothesis implies the internal entropy production is null \cite{CAMod,DeVosLibro,Hoffmann97,Durmayaz04,SalamonLibro,Wu99}, then:

\begin{equation}
\frac{\bar{J_h}}{\bar{x}}=\frac{\bar{J_c}}{\bar{y}},\label{eq:endohyp}
\end{equation}
for a linear heat transfer law:
\begin{equation}
    \bar{J_h}=\alpha\left(T_h-\bar{x}\right),\label{eq:FJh}
\end{equation}
and
\begin{equation}
    \bar{J_c}=\frac{\alpha}{\gamma}\left(\bar{y}-T_c\right),\label{eq:FJc}
\end{equation}
where $\gamma=\nicefrac{\alpha}{\beta}$ is the ratio between thermal conductances $\alpha$ and $\beta$. Under steady state operation conditions, the efficiency is written as:

\begin{equation}
    \bar{\eta}=\frac{\bar{P}}{\bar{J_h}}=1-\frac{\bar{J_c}}{\bar{J_h}}=1-\frac{\bar{y}}{\bar{x}}.\label{eq:efi}
\end{equation}

From the Eq. \ref{eq:efi} and the power output definition: $\bar{P}=\bar{J_h}-\bar{J_c}$, the heat fluxes are rewritten,

\begin{equation}
\bar{J_h}=\frac{\bar{x}}{\bar{x}-\bar{y}}\bar{P},\label{eq:NFJh}
\end{equation}
and
\begin{equation}
\bar{J_c}=\frac{\bar{y}}{\bar{x}-\bar{y}}\bar{P}.\label{eq:NFJc}
\end{equation}

Finally, by equalizing Eqs. \ref{eq:FJh}, \ref{eq:NFJh} and Eqs. \ref{eq:FJc}, \ref{eq:NFJc}, internal temperatures $\bar{x}$ and $\bar{y}$, as well as power output and the $k$--efficient power can be written as:

\begin{equation}
    \bar{x}=\frac{T_h}{1+\gamma}\left[\gamma+\frac{\tau}{1-\bar{\eta}}\right], \label{eq:xBar}
\end{equation}

\begin{equation}
    \bar{y}=\frac{T_h\left(1-\bar{\eta}\right)}{1+\gamma}\left[\gamma+\frac{\tau}{1-\bar{\eta}}\right], \label{eq:yBar}
\end{equation}

\begin{equation}
\bar{P}=\frac{\alpha\bar{\eta} T_h}{1+\gamma}\left[1-\frac{\tau}{1+\bar{\eta}}\right],\label{eq:Pbar}
\end{equation}
and
\begin{equation}
\bar{P_{\eta k}}=\bar{P_{\eta}}\bar{\eta}^{k}=\frac{\alpha\bar{\eta}^{k+1} T_h}{1+\gamma}\left[1-\frac{\tau}{1+\bar{\eta}}\right].\label{eq:Pefibar}
\end{equation}
These variables can be analyzed in some optimal operation regimens via $\bar{\eta}$. In particular, we will study the maximum $k$--efficient power regime \cite{Levario1}. For $-1\leq k$, Eq. \ref{eq:Pefibar} also yields physical achievable results. Likewise, in \cite{Levario1} Levario--Medina et al proved that the steady--state efficiency evaluated in this operating regime is:
\begin{equation}
    \bar{\eta}=\frac{2+k\left(2-\tau\right)-\sqrt{\tau}\sqrt{4+k\left(4+k\tau\right)}}{2\left(1+k\right)}.\label{eq:EtabarmaxkPEta}
\end{equation}

Therefore, by substituting Eq. \ref{eq:EtabarmaxkPEta} into Eq. \ref{eq:Pbar}, the steady--state power output at maximum $k$--efficient power results,
\begin{equation}
    \bar{P}=\frac{T_h\alpha\left[\sqrt{\tau}\left(2+k\right)-\sqrt{4+k\left(4+k\tau\right)}\right]\left[2+k\tau-\sqrt{\tau}\sqrt{4+k\left(4+k\tau\right)}\right]}{2\left(1+\gamma\right)\left[k\sqrt{\tau}-\sqrt{4+k\left(4+k\tau\right)}\right]}.\label{eq:PbarmaxkPEta}
\end{equation}

\begin{figure}
    \centering
    \includegraphics[scale=1]{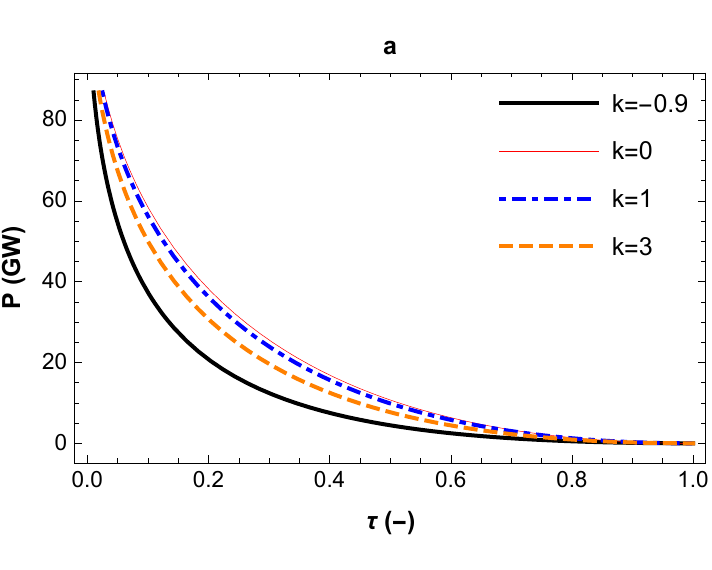}\includegraphics[scale=1]{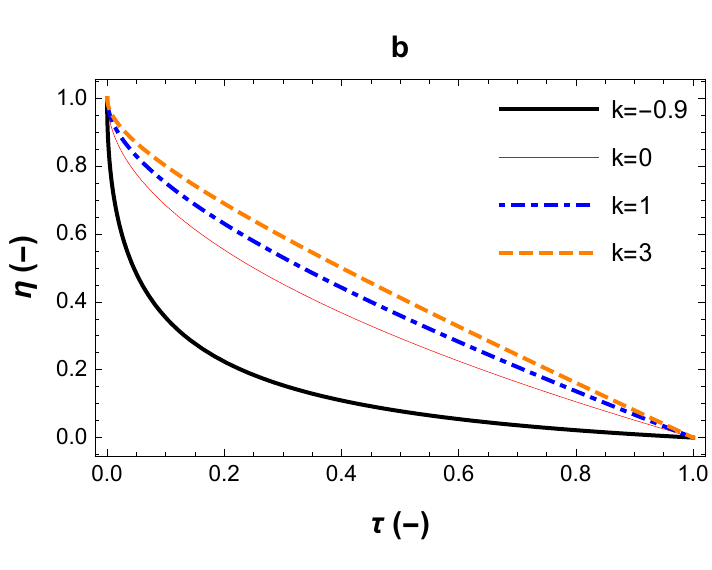}
    \caption{Curves of (a) power output and (b) efficiency, both of them evaluated at maximum $k$--efficiency power, with $k=-0.9$ (a characteristic operation mode with high dissipation), $k=0$ (maximum power output regime), $k=1$ (maximum efficient power regime) and $k=3$ (a characteristic operation mode with high efficiency) }
    \label{fig:poteficendo}
\end{figure}
In Fig. \ref{fig:poteficendo}, power output and efficiency are evaluated in the $k$--efficient power regime. Using Eqs. \ref{eq:xBar}, \ref{eq:yBar} and \ref{eq:EtabarmaxkPEta} the corresponding steady--state values of $\bar{x}$ and $\bar{y}$ as function of $T_h$ and $T_c$ at maximum $k$--efficient power regime are obtained respectively,
\begin{equation}
    \bar{x}=\frac{T_h}{1+\gamma}\left[\gamma+\frac{2\tau \left(1+k\right)}{k\tau +\sqrt{\tau}\sqrt{4+k\left(4+k\tau\right)}}\right].\label{eq:xsta}
\end{equation}
and
\begin{equation}
    \bar{y}=\frac{T_h}{1+\gamma}\left[\tau+\frac{\gamma \left\lbrace k\tau +\sqrt{\tau}\sqrt{4+k\left(4+k\tau\right)}\right\rbrace}{2\left(1+k\right)}\right].\label{eq:ysta}
\end{equation}
From previous equations for $k=0$ and $k=1$ cases, the steady--state values at maximum power output and at maximum efficient power regimes are recovered \cite{Santillan2001,Guzmanetal,Barrancoetal2}.

\section{Local stability analysis at maximum $k$--efficient power}
In this section, we present a dynamic study of local perturbations on the heat fluxes within the CA model (see Fig. \ref{fig:CAeng40}), to link the control parameter $k$, which is involved in the operation regimes, and the design parameters of the power plants (heat exchangers). It is considered $x$ and $y$  are not thermal reservoirs but real heat deposits, i. e, they represent macroscopic objects with capacity calorific $C$ \cite{Santillan2001}. Thus, the evolution of internal working temperatures change according to the following differential equations:

\begin{equation}
    \frac{dx}{dt}=\frac{1}{C} \left[ \alpha (T_h-x)-J_h \right], \label{s1}
\end{equation}
and
\begin{equation}
    \frac{dy}{dt}=\frac{1}{C}\left[J_c - \frac{\alpha}{\gamma} (y-T_c) \right]. \label{s2}
\end{equation}
As Santill\'an et al \cite{Santillan2001} emphasized that outside of the steady state but not too far, the power output of a CA heat engine depends on $x$ and $y$ in the same way that it depends on $\bar{x}$ and $\bar{y}$ at the steady state, that is, $\left(\bar{x},\bar{y}\right)$ is a fixed point, then the Taylor expansion for $P\left(x,y\right)$ is,

\begin{equation}
    P\left(x,y\right)=P\left(\bar{x},\bar{y}\right)+\left(x-\bar{x}\right)\frac{\partial P}{\partial x}+\left(y-\bar{y}\right)\frac{\partial P}{\partial y}+O\left[\left(x-\bar{x}\right)^2,\left(y-\bar{y}\right)^2\right], ... , \label{exptay}
\end{equation}
but $\left(x-\bar{x}\right)$ and $\left(y-\bar{y}\right)$ are small enough, then $P\left(x,y\right) \approx P\left(\bar{x},\bar{y}\right)$ can be assumed. Eqs. \ref{eq:NFJh} and \ref{eq:NFJc} can be rewritten for dynamic states close to steady ones as:

\begin{equation}
J_h=\frac{x}{x-y}P, \label{s3}
\end{equation}
and
\begin{equation}
J_c=\frac{y}{x-y}P. \label{s4}
\end{equation}

On the other hand, since the values of $x$ and $y$ are determined by the temperatures $T_h$ and $T_c$, in the case of maximum $k$--efficient power regime, we can express $\tau = \tau \left(\bar{x},\bar{y}\right)$ by using Eqs. \ref{eq:efi} and \ref{eq:EtabarmaxkPEta},
\begin{equation}
\tau=\frac{(1+k)\bar{y}^2}{\bar{x}(\bar{x}+k\bar{y})}. \label{s5}
\end{equation}
In a similar way to Eq. \ref{s5}, after solving Eq. \ref{eq:xBar}, we obtain an expression for $T_h$ in terms of the internal variables $\bar{x}$ and $\bar{y}$, given by

\begin{equation}
T_h=\frac{\bar{x}\left(1+\gamma\right)\left\lbrace k \tau + \sqrt{\tau \left[4+k\left(4+k \tau\right)\right]}\right\rbrace} { \left[2+k\left(2+\gamma \right)\right]\tau+\gamma \sqrt{\tau \left[4+k\left(4+k \tau \right)\right]}}. \label{s6}
\end{equation}
Finally, by substituting Eqs. \ref{s5} and \ref{s6} into Eq. \ref{eq:PbarmaxkPEta}, the steady--state power output at maximum $k$--efficient power is:
 \begin{equation}
P(\alpha, \gamma, k, \bar{x}, \bar{y}) = \frac{\alpha (\bar{x}- \bar{y})^2}{\gamma \bar{x} + (1+k+k \gamma) \bar{y}}. \label{s6a}
\end{equation}
 
Therefore, in the small perturbation approximation, we can write the dynamical equations for the temperatures $x$ and $y$ as follows:

\begin{equation}
    \frac{dx}{dt}=\frac{1}{C} \left[ \alpha (T_h-x)- \frac{x}{x-y}P(\alpha, \gamma, k, x, y) \right] \label{s7}
\end{equation}
and
\begin{equation}
    \frac{dy}{dt}=\frac{1}{C}\left[\frac{y}{x-y}P(\alpha, \gamma, k, x, y)-\frac{\alpha}{\gamma} (y-T_c) \right]. \label{s8}
\end{equation}
From the linearization technique and the fixed point stability analysis theory, we define $f(x,y)$ and $g(x,y)$ as,

\begin{equation}
    f(x,y)=\frac{\alpha}{C} \left\lbrace (T_h-x)- \frac{x}{x-y}\left[\frac{(x- y)^2}{\gamma x + (1+k+k \gamma) y}\right] \right\rbrace, \label{s9}
\end{equation}
and
\begin{equation}
    g(x,y)=\frac{\alpha}{C}\left\lbrace\frac{y}{x-y}\left[ \frac{\left(x-y\right)^2}{\gamma x+(1+k+k \gamma) y}\right]-\frac{1}{\gamma}\left(y-T_c\right) \right\rbrace. \label{s10}
\end{equation}
To analyze the system stability close to the steady state, we proceed as follows \cite{Santillan2001}: since  $\left(\bar{x},\bar{y}\right)$ is a fixed point, then $f\left(\bar{x},\bar{y}\right)=0$ and $g\left(\bar{x},\bar{y}\right)=0$. Regarding small perturbations around this fixed point, the above leads us to write $x= \bar{x}+\delta x $ and $y= \bar{y}+\delta y$, where 
$\delta x $ and $\delta y$ represent small perturbations. By expanding $f(x,y)$ and $g(x,y)$ in Taylor series around the steady state ($\bar{x},\bar{y}$), and neglecting second order terms of $\delta x$ and $\delta y$; we obtain the following matrix of linear differential equations:

\begin{eqnarray}
\left(\begin{array}{c}
  \frac{d \delta x}{dt} \\
  \frac{d \delta y}{dt}  \\
\end{array}\right)=\left(\begin{array}{cc}
  f_{x} & f_{y}  \\
  g_{x} & g_{y}  \\
\end{array}\right)
\left(\begin{array}{c}
  \delta x \\
  \delta y \\
\end{array}\right),
\label{eq:s11}
\end{eqnarray}
where $\left.f_x=\frac{\partial f}{\partial x}\right|_{\overline{x},\overline{y}}$, $\left.f_y=\frac{\partial f}{\partial y}\right|_{\overline{x},\overline{y}}$, $\left.g_x=\frac{\partial g}{\partial x}\right|_{\overline{x},\overline{y}}$, $\left.g_y=\frac{\partial g}{\partial y}\right|_{\overline{x},\overline{y}}$. From Eqs. \ref{s9} and \ref{s10} and by using Eqs. \ref{eq:xsta} and \ref{eq:ysta} we get:

\begin{equation}
    f_x=-\frac{2\alpha(1+\gamma)}{C}\left\lbrace\frac{2(1+k)\left[a\left(1+k+k\gamma\right)+\gamma +k\gamma \right] + k\left[4+k(4+a)\right](1+k+k\gamma)\tau + k^3\tau^2(1+k+k\gamma)}{\left[2(1+k)\gamma +a(1+k+k\gamma)+k(1+k+k\gamma)\right]^2}\right\rbrace, \label{eq:dparfx}
\end{equation}

\begin{equation}
    f_y=\frac{\alpha(1+\gamma)}{C}\left\lbrace\frac{4(1+k)^3}{\left[(1+k)(k\tau +a) + \gamma (2+2k+ak+k^2\tau)\right]^2}\right\rbrace, \label{eq:dparfy}
\end{equation}

\begin{equation}
    g_x=\frac{\alpha(1+\gamma)}{C}\left\lbrace\frac{(1+k)(k\tau+a)^2}{\left[(1+k)(k\tau +a) + \gamma (2+2k+ak+k^2\tau)\right]^2}\right\rbrace, \label{eq:dpargx}
\end{equation}
and
\begin{equation}
    g_y=-\frac{\alpha(1+\gamma)}{C \gamma}\left\lbrace\frac{(1+k)(k\tau+a) \left[\gamma (1+k)(k\tau +a) (4+4k+ak+k^2\tau)\right]}{\left[(1+k)(k\tau +a) + \gamma (2+2k+ak+k^2\tau)\right]^2}\right\rbrace, \label{eq:dpargy}
\end{equation}
where $a=\sqrt{\tau\left[4+k(4+k\tau)\right]}$. Let  $\lambda_1$ and $\lambda_2$ be the eigenvalues of the Jacobian matrix given by the first term on the right--hand of Eq. \ref{eq:s11}, then the temporal evolution for this equation system is $\delta \vec{r} = e^{\lambda t}\vec{u}$, with $\delta \vec{r}=(\delta x,\delta y)$ and $\vec{u}=(u_x,u_y)$.
Therefore, the eigenvalues $\lambda_1$ and $\lambda_2$ can be calculated by means of the characteristic equation,

\begin{equation}
   |A-\lambda I|=\left(f_x - \lambda \right) \left(g_x - \lambda \right) - f_x g_y = 0,  \label{eq:poly}
\end{equation}
with $A$ the Jacobian matrix. After solving Eq. \ref{eq:poly}, it is shown that both eigenvalues $\lambda_1$ and $\lambda_2$ are function of $\gamma$, $C$, $x(\tau, \gamma, k)$, $y(\tau, \gamma, k)$ and $k$ parameters,

\begin{equation}
    \lambda_{1}=-\frac{\alpha(1+\gamma)}{2C\gamma}\left\{ \frac{x^{2}\gamma^{2}+y^{2}\left(1+k+k\gamma\right)^{2}+2xy\gamma\left(2+2k+k\gamma\right)+\left(1+k\right)^{2}y^{2}+2yb\gamma\left(1+k\right)+\gamma^{2}\left(x+ky\right)^{2}}{\left[x\gamma+y\left(1+k+k\gamma\right)\right]^{2}}\right\},\label{lamb1}
\end{equation}
and
\begin{equation}
    \lambda_{2}=-\frac{\alpha(1+\gamma)}{2C\gamma}\left\{ \frac{x^{2}\gamma^{2}+y^{2}\left(1+k+k\gamma\right)^{2}+2xy\gamma\left(2+2k+k\gamma\right)-\left[\left(1+k\right)^{2}y^{2}+2yb\gamma\left(1+k\right)+\gamma^{2}\left(x+ky\right)^{2}\right]}{\left[x\gamma+y\left(1+k+k\gamma\right)\right]^{2}}\right\},\label{lamb2}
\end{equation}
\begin{figure}
    \centering
    \includegraphics[scale=1]{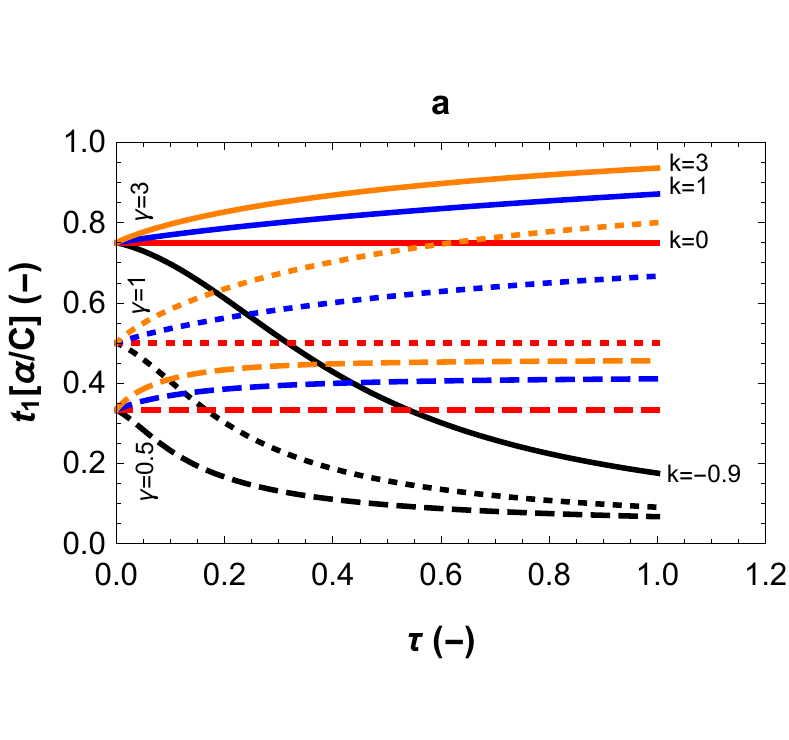}\includegraphics[scale=1]{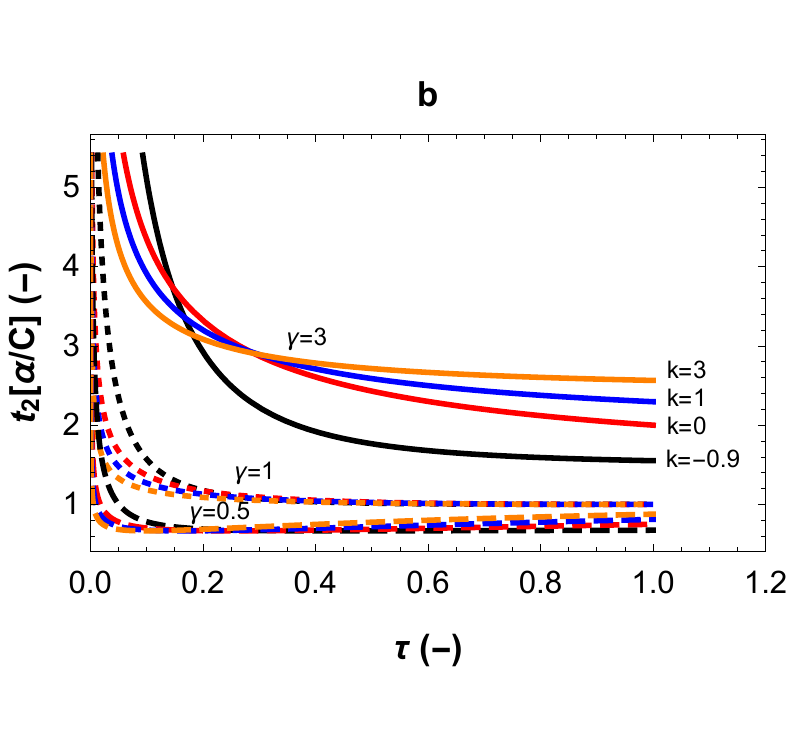}
    \caption{Plots of the normalized relaxation times $\nicefrac{t_1 \alpha}{C}$ and $\nicefrac{t_2 \alpha}{C}$ versus $\tau$ for several values of $k$ at maximum $k$--efficient power regime. In a) for three different values of $\gamma$, $t_1$ increases asymptotically while $k\geq 0$, otherwise its behavior decreases. In b) for the same values of $\gamma$, $t_2$ always decays exponentially.}
    \label{fig:timheng}
\end{figure}
where $x=x(\tau, \gamma, k)$ and $y=y(\tau, \gamma, k)$ are given by Eqs. \ref{eq:xsta} and \ref{eq:ysta} respectively and $b=\left(x^{2}-2kxy-k^{2}y^{2}\right)$. As both of them eigenvalues are real and negative, then  $\delta x$ and $\delta y$ perturbations monotonically converge to the steady state of the system, which is stable. In addition, the relaxation times are defined as $t_1 = \nicefrac{1}{|\lambda_1|}$ and $t_2 = \nicefrac{1}{|\lambda_2|}$, they are plotted versus $\tau$ for different values of the $k$ parameter and for three different values of the $\gamma$ parameter (see Fig. \ref{fig:timheng}). We observe for all the interval $0<\tau <1$, the steady state is stable because any perturbation would decay exponentially. The cases that correspond to the maximum power conditions ($\gamma =1$ and $k=0$) \cite{Santillan2001,Guzmanetal}, and maximum efficient power regime ($\gamma =1$ and $k=1$) \cite{Barrancoetal2} are also shown in Fig. \ref{fig:timheng}. The relaxation times increase when $\gamma$ increases. However, when $\gamma>1$, $t_2 \rightarrow \infty$ the stability is lost. In the opposite case $\gamma <0$, $t_2$ reveals a minimum, which strongly depends on the $\gamma$ and $k$ values.

The general solution of the dynamic system (Eq. \ref{eq:s11}) is,

\begin{equation}
    \delta \vec{r}=G_1e^{\lambda_1t}\vec{u}_1+G_2e^{\lambda_2t}\vec{u}_2,\label{gensol}
\end{equation}
where $G_1$ and $G_2$ are constants that fulfill initial conditions for Eq. \ref{gensol}, $\vec{u}_1$ and $\vec{u}_2$ are the eigenvectors belong to $\lambda_1$ and $\lambda_2$, respectively. Thus, the eigenvectors that characterize asymptotic stability are:

\begin{equation}
    \vec{u}_{1}=\left(1,\frac{x^{2}\gamma^{2}-y^{2}\left(1+k\right)^{2}+ky\gamma^{2}\left(2x+ky\right)-\left[\left(1+k\right)^{2}y^{2}+2yb\gamma\left(1+k\right)+\gamma^{2}\left(x+ky\right)^{2}\right]}{2\gamma x^{2}\left(1+k\right)}\right), \label{eivect1}
\end{equation}
and
\begin{equation}
    \vec{u}_{2}=\left(1,\frac{x^{2}\gamma^{2}-y^{2}\left(1+k\right)^{2}+ky\gamma^{2}\left(2x+ky\right)+\left(1+k\right)^{2}y^{2}+2yb\gamma\left(1+k\right)+\gamma^{2}\left(x+ky\right)^{2}}{2\gamma x^{2}\left(1+k\right)}\right).\label{eivect2}
\end{equation}

\begin{figure}
    \centering
    \includegraphics[scale=1]{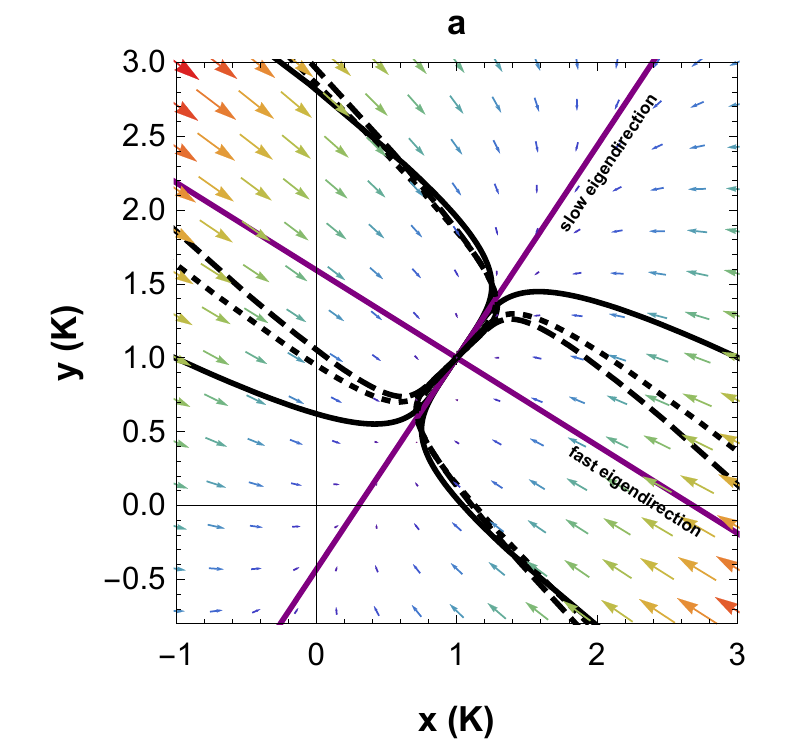}\includegraphics[scale=1]{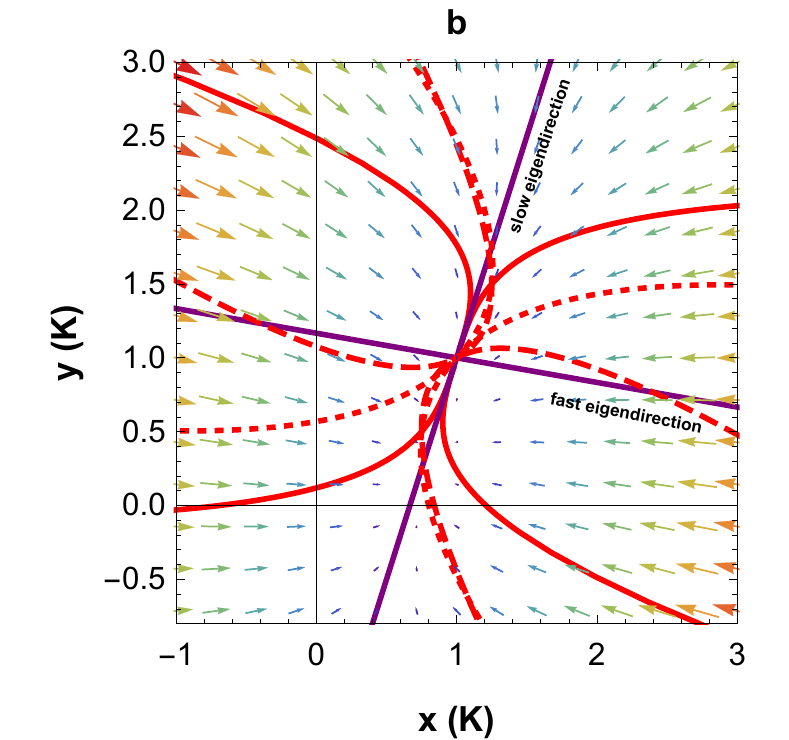}\newline
    \includegraphics[scale=1]{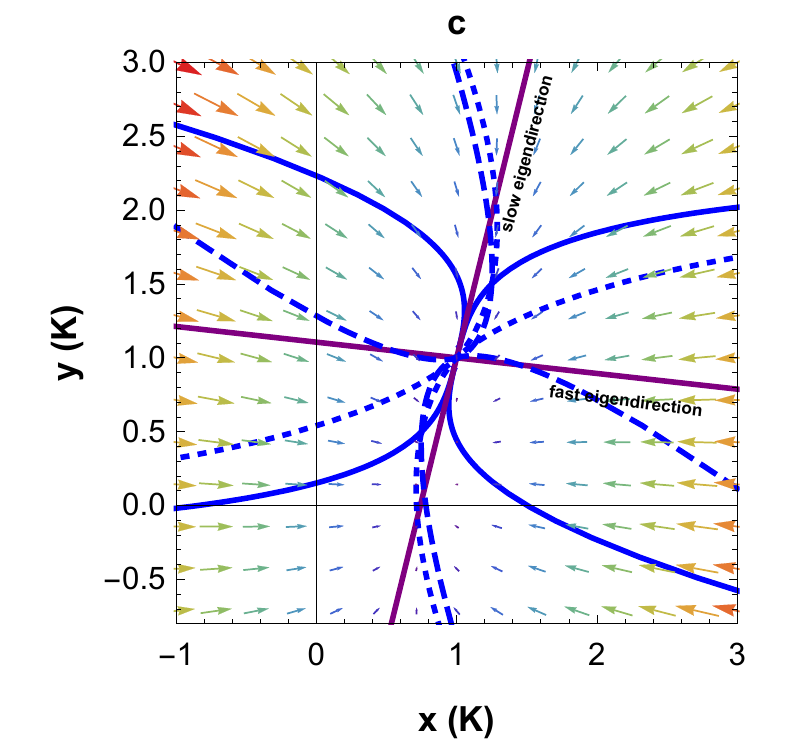}\includegraphics[scale=1]{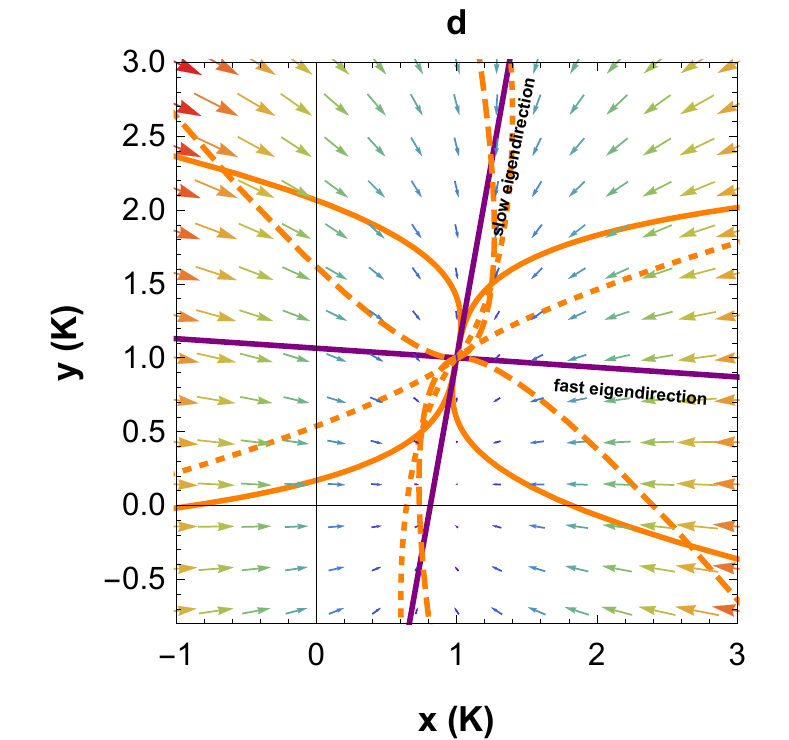}
    \caption{Qualitative phase portrait of $x(t)$ vs $y(t)$ for $\tau =0.5$, $\gamma =0.5$ (dashed line), $\gamma =1$ (dotted line) and $\gamma =3$ (solid line). In a) ($k=-0.9$, black), b) ($k=0$, red), c) ($k=1$, blue) and d) ($k=3$, orange) trajectories approach to $\left(\bar{x},\bar{y}\right)$ (vector field), according to features of the eigenvectors (fast and slow directions). Both of the eigenvalues are negative and [$x(t)$,$y(t)$] decay exponentially to the steady state.}
    \label{fig:spheng}
\end{figure}
As can be noted in Fig. \ref{fig:timheng}, for all the considered parameters the inequality $0<t_1<t_2$ is fulfilled. That is, the ratio $\nicefrac{t_2}{t_1}>1$ allows us to identify the fast eigendirection ($\vec{u}_1$) and the slow eigendirection ($\vec{u}_2$) for a given value of $\gamma$, both eigendirections reflect the dynamic preference of the system to reach a steady state. Then the dynamic behaviour of the working temperatures is represented by means of a phase space portrait. In Fig. \ref{fig:spheng} the qualitative phase portrait for a heat engine is shown for the same two values of the $\gamma$ parameter and all cases ($k=-0.9$, $k=0$, $k=1$ and $k=3$) considered in the eigenvalues behaviour (see Fig. \ref{fig:timheng}). We can see in all cases a), b), c) and d) of Fig.\ref{fig:spheng}, the trajectories converge to steady state, tangent to the slow eigendirection and parallel to the fast one. In addition, for $\gamma<0$ the ratio $e^{\left(\lambda_1 -\lambda_2 \right)t_0}<1$ when $k\rightarrow -1$; this means that the working temperatures, under the same initial conditions, will converge fast to the stable point. For $\gamma > 0$, $e^{\left(\lambda_1 -\lambda_2 \right)t_0}>1$ when $k\rightarrow \infty$, i.e, the working temperatures, by contrast, will converge slowly \cite{Jeffreys62}.

Since the eigenvectors $\vec{u}_1$ and $\vec{u}_2$ depend strongly on each operation mode characterized by a $k$ parameter, there is a relation that represents the change of the eigendirections between two different $k$--efficient power regimes $\left(k_1, k_2\right)$. This change is manifested by a phase portrait rotation. The rotation angles for the fast ($\theta_{fev}$) and slow ($\theta_{sev}$) eigenvectors are given respectively by,

\begin{equation}
    \theta_{fev}=\arccos\left[\left\langle \frac{\vec{u}_{1}^{\left(k_{1}\right)}}{\left\Vert \vec{u}_{1}^{\left(k_{1}\right)}\right\Vert },\frac{\vec{u}_{1}^{\left(k_{2}\right)}}{\left\Vert \vec{u}_{1}^{\left(k_{2}\right)}\right\Vert }\right\rangle \right], \label{angev1}
\end{equation}
and
\begin{equation}
   \theta_{sev}=\arccos\left[\left\langle \frac{\vec{u}_{2}^{\left(k_{1}\right)}}{\left\Vert \vec{u}_{2}^{\left(k_{1}\right)}\right\Vert },\frac{\vec{u}_{2}^{\left(k_{2}\right)}}{\left\Vert \vec{u}_{2}^{\left(k_{2}\right)}\right\Vert }\right\rangle \right], \label{angev2}
\end{equation}
where $\left\langle \cdot,\cdot\right\rangle$ is the inner product of the unit vectors $\hat{u}_{1,2}^{(k_1)}$ and $\hat{u}_{1,2}^{(k_2)}$. In a recent work \cite{Levario1}, by means of a generalization of the efficient power regime, the restructuring conditions for some power plants were established, i. e, they may stop working in their configuration space with low efficiency to operate within a high efficiency and low dissipation zone. The restructuring conditions allow to obtain the best values of the control parameters ($\alpha$, $\gamma$ and $\tau$) to provide a better performance of the power plants. In the following subsection we apply the previous local stability theory to real plants that can be reconfigured energetically \cite{Levario1}.

\subsection{Effects of local stability for some power plants}
The study of thermal perturbations is not exclusive to certain types of heat engines, in which the design directly affects the intrinsic cyclic variability of the working fluid \cite{Rochaetal2002,Curtoetal2011}. In fact, energy converters whose main objective is generating a specific type of energy by means of a primary source (power plants) contain a large number of mechanical couplings, that externally disturb their operation when these power plants are operating within a particular operation regime. The influence of the operation modes (characterized by the $k$ parameter) on some control parameters, particularly the thermal conductances, will be reflected in the quickness of convergence to the respective steady states.

\begin{table}
\begin{centering}
\begin{tabular}{|c|c|c|c|c|c|c|c|c|c|c|}
\hline 
Power Plant & \multicolumn{5}{c|}{Original Configuration (OC)} & \multicolumn{5}{c|}{Restructured Configuration (RC)}\tabularnewline
\hline 
Larderello & $T_{h}\,\left[\textrm{K}\right]$ & $T_{c}\,\left[\textrm{K}\right]$ & $\tau\,\left[-\right]$ & $k\,\left[-\right]$ & $\gamma\,\left[-\right]$ & $T_{h}\,\left[\textrm{K}\right]$ & $T_{c}\,\left[\textrm{K}\right]$ & $\tau\,\left[-\right]$ & $k\,\left[-\right]$ & $\gamma\,\left[-\right]$\tabularnewline
(Italy, 64) & $523$ & $353$ & $0.675$ & $-0.2211$ & $3$ & $518.032$ & $353$ & $0.6814$ & $0.2838$ & $2.798$\tabularnewline
\hline 
Toshiba & $T_{h}\,\left[\textrm{K}\right]$ & $T_{c}\,\left[\textrm{K}\right]$ & $\tau\,\left[-\right]$ & $k\,\left[-\right]$ & $\gamma\,\left[-\right]$ & $T_{h}\,\left[\textrm{K}\right]$ & $T_{c}\,\left[\textrm{K}\right]$ & $\tau\,\left[-\right]$ & $k\,\left[-\right]$ & $\gamma\,\left[-\right]$\tabularnewline
(109FA, 04) & $1573$ & $303$ & $0.193$ & $-0.4569$ & $3$ & $1506.150$ & $303$ & $0.2012$ & $0.8412$ & $2.807$\tabularnewline
\hline 
West Thurrock & $T_{h}\,\left[\textrm{K}\right]$ & $T_{c}\,\left[\textrm{K}\right]$ & $\tau\,\left[-\right]$ & $k\,\left[-\right]$ & $\gamma\,\left[-\right]$ & $T_{h}\,\left[\textrm{K}\right]$ & $T_{c}\,\left[\textrm{K}\right]$ & $\tau\,\left[-\right]$ & $k\,\left[-\right]$ & $\gamma\,\left[-\right]$\tabularnewline
(UK, 62) & $838$ & $298$ & $0.356$ & $-0.2966$ & $3$ & $818.558$ & $298$ & $0.3641$ & $0.4217$ & $2.831$\tabularnewline
\hline 
\end{tabular}
\par\end{centering}
\caption{\label{tab:heatdata}Some operation parameters associated with heat flow $(T_h,T_c,\tau)$ and the performance regime $(k)$ for one combined cycle (Toshiba) and two simple--cycle (Larderello and West Thurrock) power plants.}
\end{table}
By taking as examples some power plants (West Thurrock, Larderello and Toshiba) whose performance is not located in the so-called optimal operation zone (high efficiency and low dissipation) \cite{Levario1}, we study the role that $k$ parameter plays when systems have been locally disrupted. Table \ref{tab:heatdata} shows some reported parameters for the aforementioned power plants during their operation (Original Configuration), and those theoretically calculated so that they can work with the same power output but with greater efficiency (Restructured Configuration) \cite{Levario1}.

\begin{figure}
    \centering
    \includegraphics[scale=0.8]{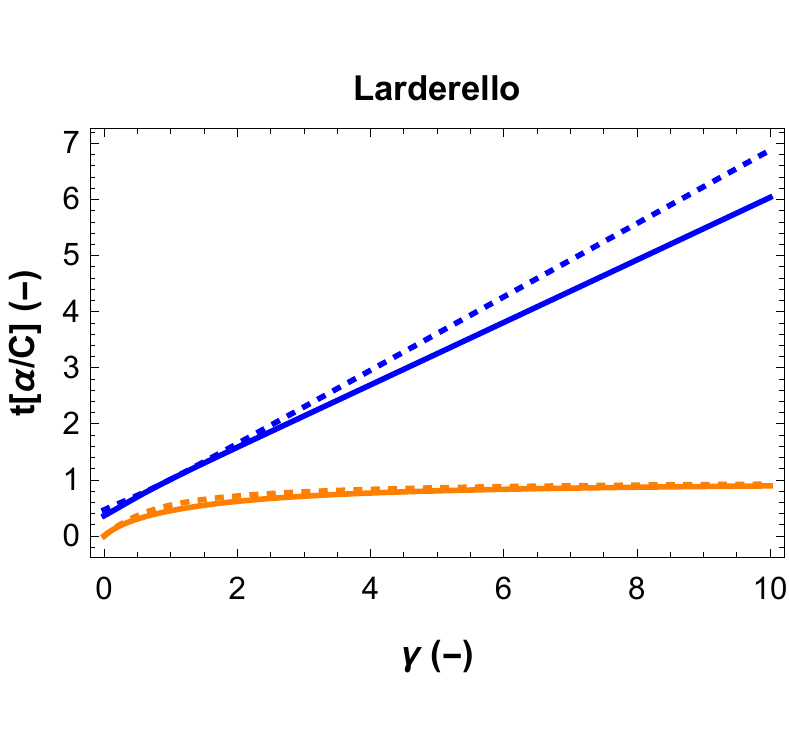}\includegraphics[scale=0.8]{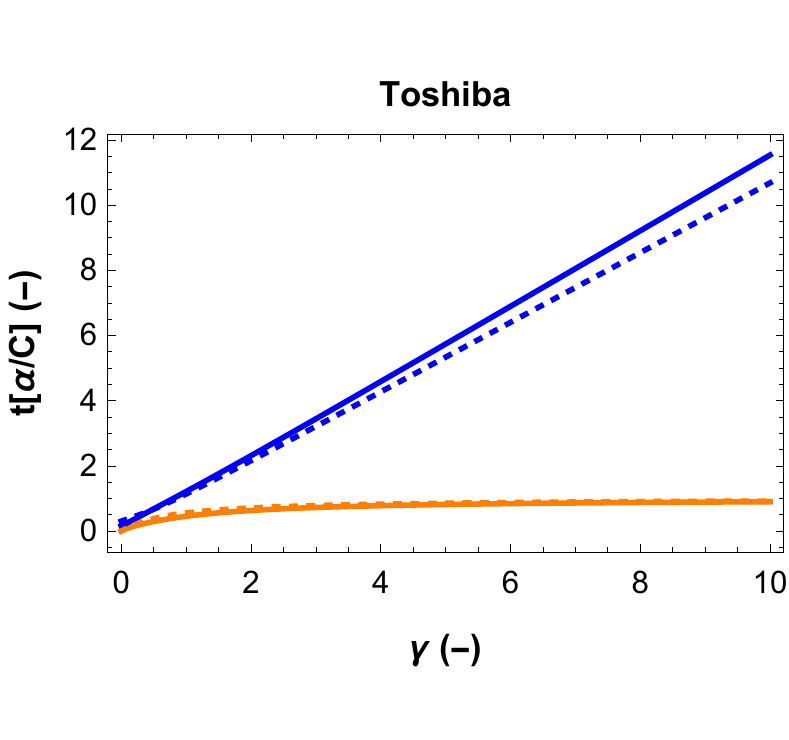}\includegraphics[scale=0.8]{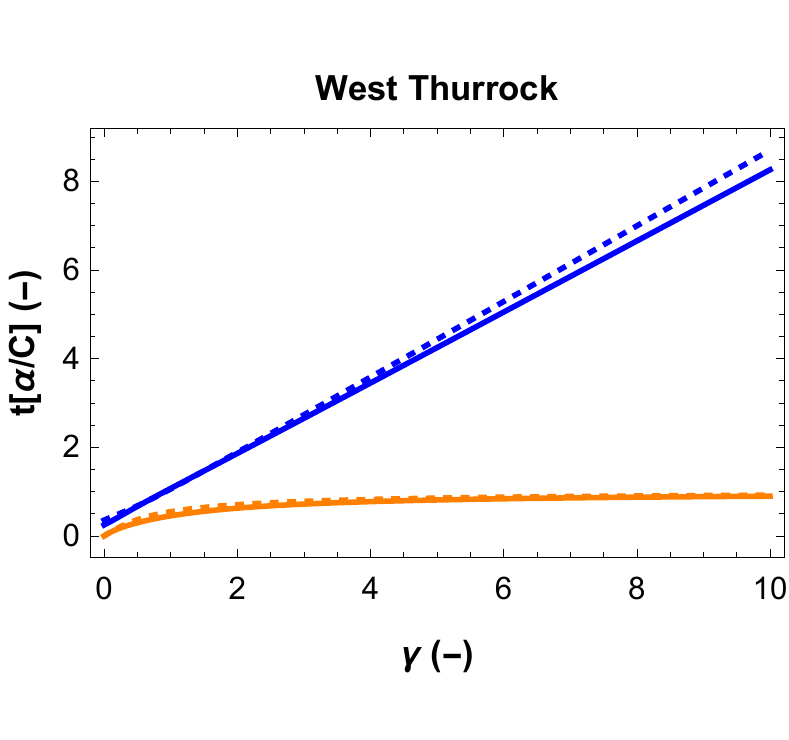}
    \caption{Graphs of the normalized relaxation times $\left(\nicefrac{\alpha t_{1}^{OC}}{C},\nicefrac{\alpha t_{1}^{RC}}{C}\right)$ (in blue) and $\left(\nicefrac{\alpha t_{2}^{OC}}{C},\nicefrac{\alpha t_{2}^{RC}}{C}\right)$ (in orange) versus $\gamma$ for the power plants reported in Table \ref{tab:heatdata}. The superscript $OC$ (solid line) corresponds to the original configuration of the plant while the one with $RC$ (dashed line) is associated to a restructured configuration.}
    \label{fig:pwpltim}
\end{figure}
The asymptotic stability, which power plants can experience when thermal perturbations are taken into account, is visualized through relaxation times (see Fig. \ref{fig:pwpltim}). Normalized relaxation times $\nicefrac{\alpha t_{1}^{OC}}{C}$ and $\nicefrac{\alpha t_{2}^{OC}}{C}$ correspond to the operation of the power plants in their original configuration, while $\nicefrac{\alpha t_{1}^{RC}}{C}$ and $\nicefrac{\alpha t_{2}^{RC}}{C}$ match the one they could have in the restructured configuration. In Fig. \ref{fig:pwpltim} after a small perturbation, it is observed that Larderello and West Thurrock plants reach the steady state in a shorter time when they are operating in the original regimes, that is, $t_{1,2}^{OC}>t_{1,2}^{RC}$. In case of Toshiba plant $t_1^{RC} \approx t_1^{OC}$ and $t_2^{RC}>t_2^{OC}$ for $\gamma>1$.

\begin{figure}
    \centering
    \includegraphics[scale=0.8]{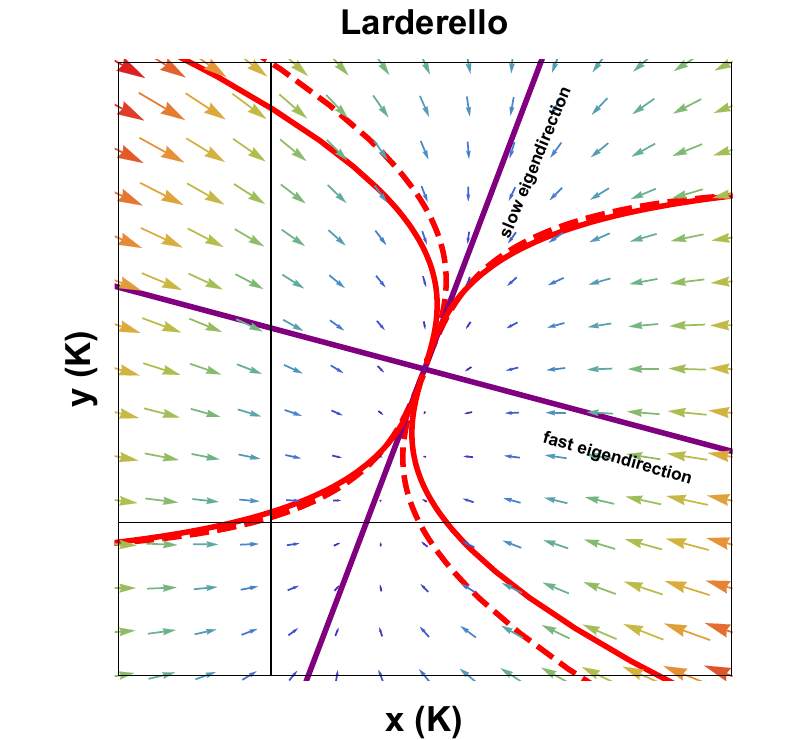}\includegraphics[scale=0.8]{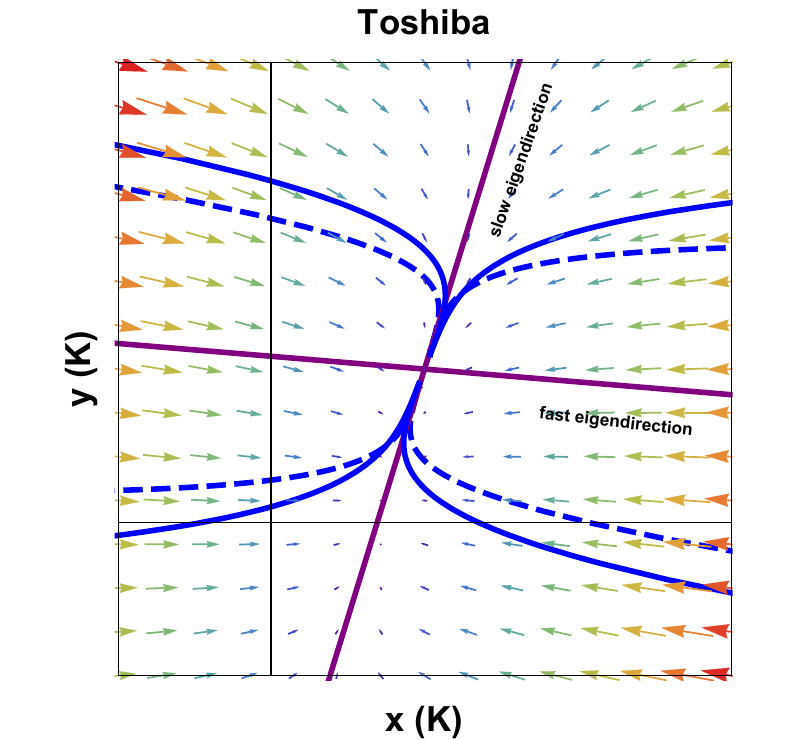}\includegraphics[scale=0.8]{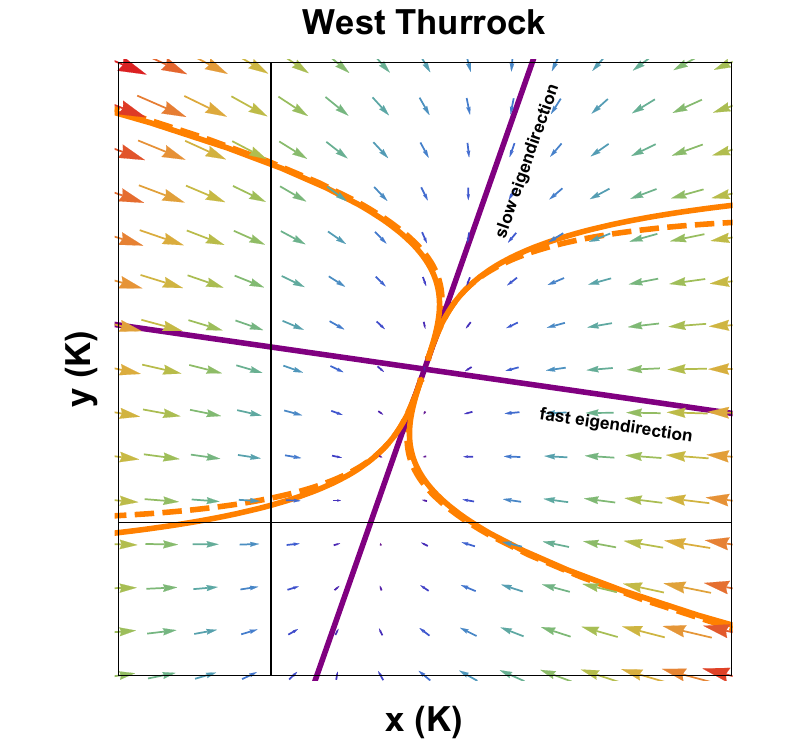}
    \caption{Qualitative phase portrait of $x(t)$ and $y(t)$ for Larderello plant (in red), Toshiba plant (in blue) and West Thurrock plant (in orange). In all the phase portraits, the evolution of the disturbed working temperatures is represented when the plants are operating in their original configuration (dashed lines), and in some restructured configuration (solid lines), the parameters used are in Table \ref{tab:heatdata}.}
    \label{fig:pwpltspha}
\end{figure}
The dynamic behavior of thermal perturbations, during the operation of the above power plants, can be represented through their respective phase portraits (see Fig. \ref{fig:pwpltspha}). The speed of convergence towards the steady state of the restructuring configuration with respect to the original one, can be analyzed via the asymptotic behavior \cite{Jeffreys62} of $f(x,y)$ for $k_{RC}$ and $k_{OC}$,

\begin{equation}
    L_x=\begin{array}{c}\lim\\\left(x,y\right)\rightarrow\left(\bar{x},\bar{y}\right)\end{array}\left[\frac{f_{k_{RC}}\left(x,y\right)}{f_{k_{OC}}\left(x,y\right)}\right]. \label{eq:limasym1}
\end{equation}

This limit is related in some way to the fast eigendirection within the space portraits. Thus, for Larderello plant: $L_x=0.994$, Toshiba plant: $L_x=1.072$ and West Thurrock plant: $L_x=0.976$. In case of $g(x,y)$ function, its asymptotic demeanor is:

\begin{equation}
    L_y=\begin{array}{c}\lim\\\left(x,y\right)\rightarrow\left(\bar{x},\bar{y}\right)\end{array}\left[\frac{g_{k_{RC}}\left(x,y\right)}{g_{k_{OC}}\left(x,y\right)}\right]. \label{eq:limasym2}
\end{equation}

While this limit is linked to the slow eigendirection, for Larderello plant: $L_y=1.072$, Toshiba plant: $L_y=0.957$ and West Thurrock plant: $L_y=1.060$. The above results, as well as the analysis of relaxation times for power plants show the effect of the restructuring configurations for the study of small thermal perturbations.

In the following section, we analyze the effect of the generalization parameter $k$ on a parameter (Lyapunov function) that guarantees the equilibrium point (steady state) as an attractor for any evolution path of the thermal perturbation during the operation of a heat engine.

\section{Global stability analysis at maximum $k$--efficient power}
In the previous section, we studied the local stability of a CA--heat engine operating at maximum $k$--efficient power. In particular for power plants that are currently operating \cite{Levario1,enviadoJNET}. In order to analyze the global asymptotic behavior of a dynamic system, a suitable Lyapunov function needs to be found to show all the trajectories of thermal perturbations of a heat engine converge to the equilibrium point. There are direct and indirect methods to find a Lyapunov function \cite{Tzafestas14,Khalil15} that enables the stability study of a system. Therefore, in this Section we apply a direct method which consist in finding a Lyapunov function that guarantees the global asymptotic behavior of the heat engine model shown in Fig. \ref{fig:CAeng40}, by means of Lyapunov stability theory. For autonomous systems of the form $\nicefrac{dx}{dt}=h(x)$ \cite{I1,I2,Tzafestas14,Khalil15}, the candidate Lyapunov function must satisfy:

\begin{description}

  \item[\textit{\bf (i)}] $V(\bar{x},\bar{y})=0$,
  \item[\textit{\bf (ii)}] $V(x,y)$ must be positive definite in a region around the steady state,
  \item[\textit{\bf(iii)}] $V(x,y)$ must be radially unbounded, 
  \item[\textit{\bf (iv)}] $ \dot{V}(x,y)$ must be negative definite for the same region around the steady state.
 
\end{description}

The Lyapunov’s direct method based on the Krasovskii's theorem requires constructing the Lyapunov function. In accordance with the procedure of \cite{I1,I2} (Krasovskii's method), we find a symmetric matrix of the form $J(X)=A(X)+A^{T}(X)$; with $X=(x,y)$, $A$ is the Jacobian matrix of $\delta \vec{r}$ (Eq. \ref{eq:s11}) and $A^T$ represents its transpose matrix. Thus, $J(X)$ is:

\begin{eqnarray}
J=\frac{\alpha(1+ \gamma)}{C} \left[\begin{array}{cc}
 J_{xx} (\gamma,\tau,C,k) &  J_{xy} (\gamma,\tau,C,k) \\
 J_{yx} (\gamma,\tau,C,k) &  J_{yy} (\gamma,\tau,C,k) 
\end{array}\right],
 \label{mjac}
\end{eqnarray}
where,

\begin{equation}
J_{xx} (\gamma,\tau,C,k) = -\frac{8(1+k)(a+ak+\gamma +4k \gamma + ak \gamma) + k\tau \left[4+(4+a)k\right](1+k+k\gamma)+ 4 k^3\tau^2(1+k+k\gamma) }{\left[2\gamma(1+k) +a(1+k+k\gamma)+k\tau(1+k+k\gamma) \right]^2}, \label{J11}
\end{equation} 

\begin{equation}
J_{xy} (\gamma,\tau,C,k)= J_{yx} (\gamma,\tau,C,k)= \frac{(1+k)\left[ 4(1+k)^2 + (a+k\tau)^2 \right]}{\left[2\gamma(1+k) +a(1+k+k\gamma)+k\tau(1+k+k\gamma) \right]^2} \label{J12}
\end{equation}
and
\begin{equation}
J_{yy} (\gamma,\tau,C,k)= - \frac{(1+k)(a+k\tau)\left\lbrace (1+k)(a+k\tau) + \gamma \left[4 + k (4+a+k\tau)\right] \right\rbrace }{\gamma \left\lbrace(1+k)(a+k\tau) + \gamma \left[2+k(2+a+k\tau)\right]   \right\rbrace^2}. \label{J22}
\end{equation} 

The candidate Lyapunov function at maximum $k$-efficient power regime is given by \cite{Tzafestas14,Khalil15},
\begin{equation}
V(X) = \left<F(X),F(X)\right> = \left|F(X)\right|^2, \label{Lfunc}
\end{equation} 
where each element in vector $F(X)=\left[f(x,y),g(x,y)\right]$ is given by Eqs. \ref{s9} and \ref{s10}, respectively. Then,

\begin{equation}
\hspace{-0.2cm}V(x,y)=\frac{\alpha ^2}{C^2}\left\lbrace\frac{\gamma^2 \left[(1+k)T_h y-(x+ky)(x-T_h \gamma +x \gamma)\right]^2+\left[y^2(1+k)(1+\gamma)-T_c(x\gamma+y+yk+yk\gamma)\right]^2}{\gamma^2 \left[\gamma x + y(1+k+k\gamma)\right]^2} \right\rbrace. \label{lyapf}
\end{equation}

\begin{figure}
    \centering
    \includegraphics[scale=0.9]{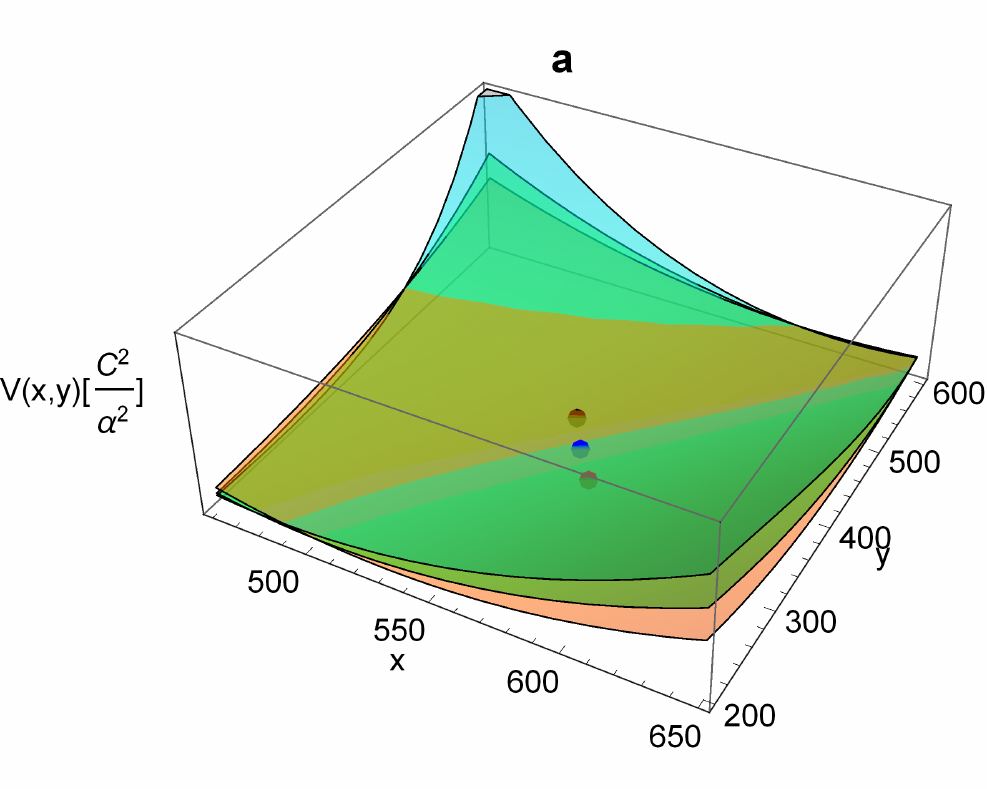}\includegraphics[scale=0.9]{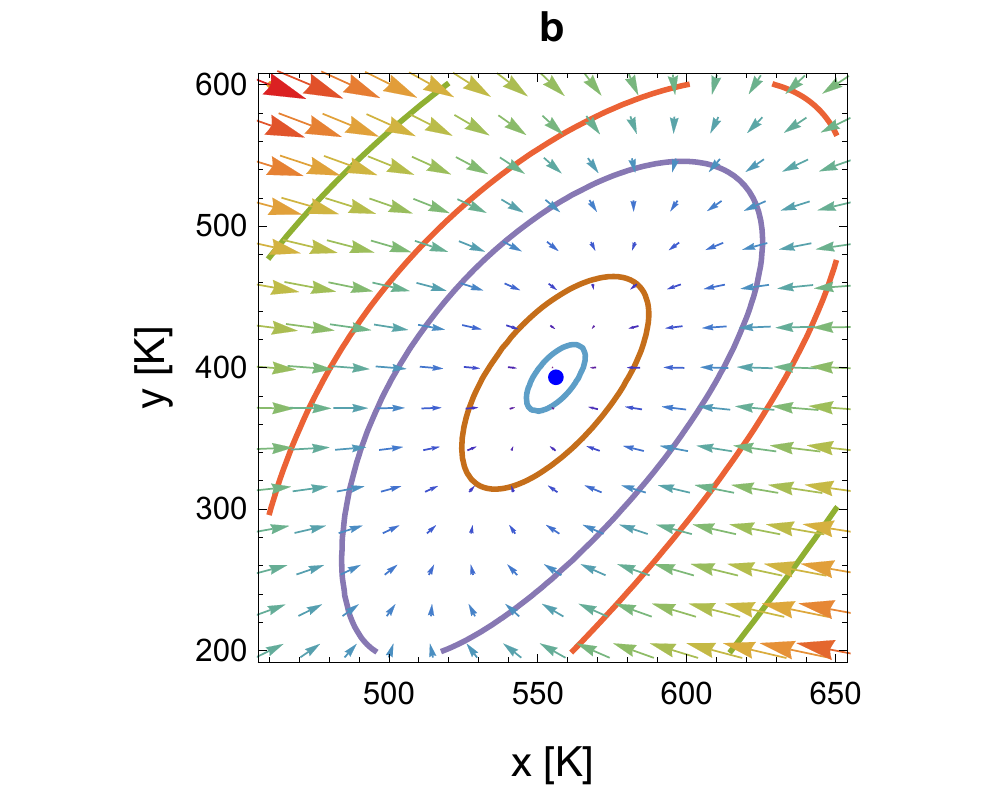}
    \caption{Effects of the $k$ parameter on the Lyapunov method. In a) three different normalized Lyapunov surface plots $\left[V(x,y)\right] \left(\nicefrac{C^2}{\alpha^2}\right)$ for $k=-0.5$ (blue surface), $k=0$ (green surface) and $k=1$ (orange surface) are shown; black, blue and red points are the limit cycles for each surface. In b) the qualitative plot of the level curve corresponding to $k=0$ is depicted, the associated vector field shows how the steady state (blue point) is global asymptotically stable for any thermal perturbation. We considered the values of $\gamma=3$, $T_h=600 K$ and $T_c=300 K$.}
    \label{fig:LyF1}
\end{figure}

Thus, we can verify, on the one hand, the function $V(x,y)$ satisfies $V(\bar{x},\bar{y})=0$; that is, the steady-state regime is the limit cycle of $V(x,y)$ and, on the other, $V(x,y)$ is unbounded because $V(x,y)\rightarrow \infty$ when $(x,y)\rightarrow \infty$. Additionally, the \textbf{(iv)} condition can be written as

\begin{equation}
    \dot V(x,y)= \frac{\partial V}{\partial x} \dot x + \frac{\partial V}{\partial y} \dot y = \langle \nabla V,\dot X \rangle < 0, \label{eq:ivcond}
\end{equation}
where $\nicefrac{\partial V}{\partial x}$ and $\nicefrac{\partial V}{\partial y}$ are the directional derivatives along the $x$ and $y$ axes respectively; both derivatives are analytic expressions. For all values of $k$ it is fulfilled that $\dot{V}(x,y)<0$ (see Fig. \ref{fig:LyF1}), which means the trajectories move towards surfaces with $V(x,y)<\Omega$. For instance, if the projections of the heat flux vector $\dot{X}$ on the normal $\nabla V$, are equal to zero, the trajectories lie on the surface $V(x,y)=\Omega$. As mentioned, each physically accessible operation mode for a heat engine is represented by only one value of $k$. Therefore, when the global asymptotic stability conditions are satisfied for an operation mode, the system will converge to a different limit cycle (steady state) as depicted on the Lyapunov surfaces of Fig. \ref{fig:LyF1} b.

In the following subsection, we study the restructuring condition effects for the power plants presented above, with the aim of finding the best speed of convergence at each limit cycle when the system lies to an inner Lyapunov surface with a smaller value.

\subsection{Effects of global stability for some power plants}
\begin{figure}
    \centering
    \includegraphics[scale=0.9]{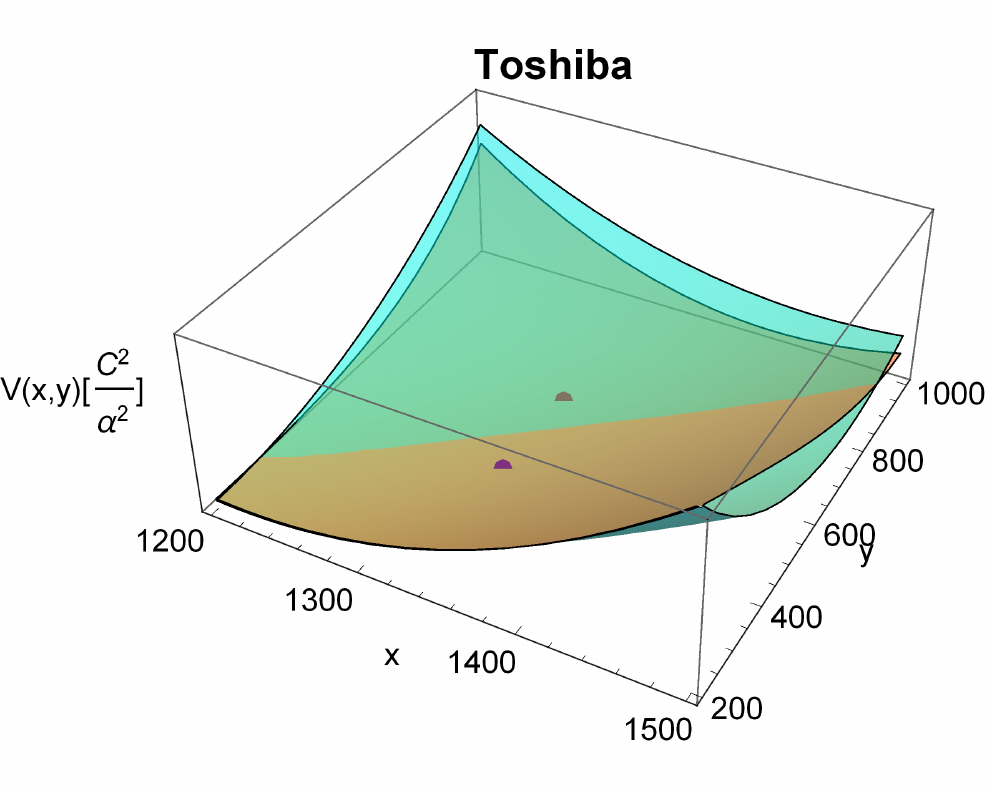}\includegraphics[scale=0.9]{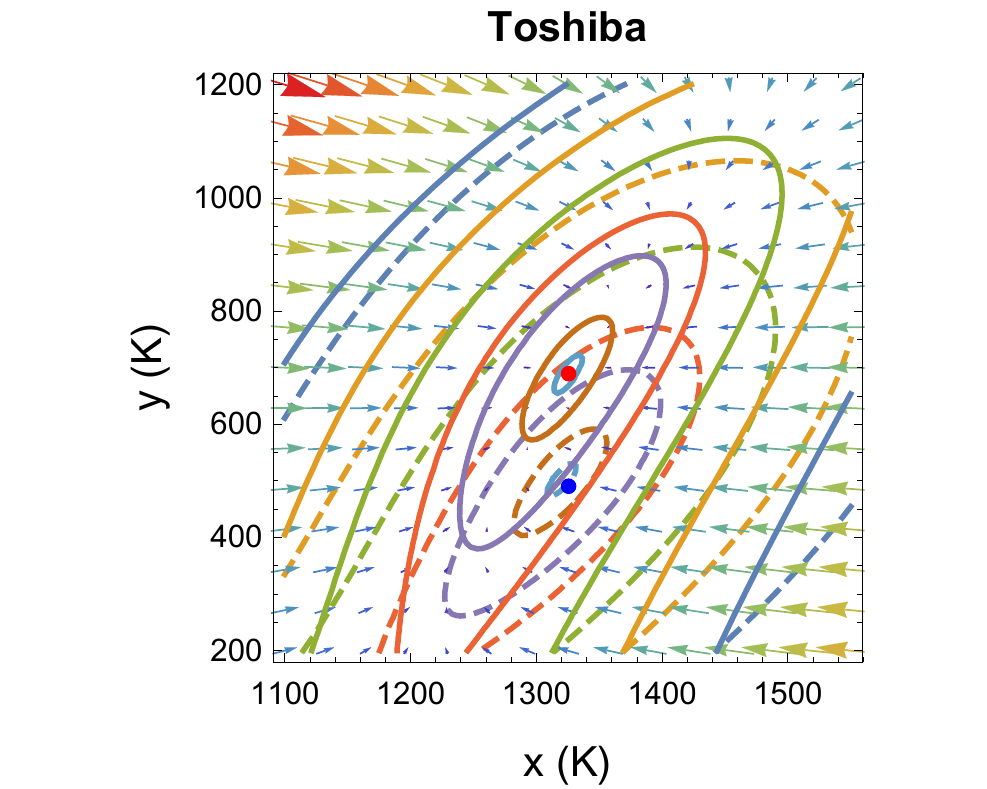}
    \caption{Qualitative plots of normalized Lyapunov surfaces and its level curves for Toshiba power plant in its original (solid curves) and restructured (dashed curves) configurations. The associated vector field shows that the steady states (blue and red points) are global asymptotically stables. All of the parameter values are on Table \ref{tab:heatdata}.}
    \label{fig:Toshlyap}
\end{figure}

In recent works \cite{Levario1,enviadoJNET}, the authors established conditions so that power plants can be operated from a zone with low efficiency (LE) and high dissipation (HD) into another one with high efficiency (HE) and low dissipation (LD), through an energetic restructuring process. In a similar way to subsection 3.1, we constructed some Lyapunov functions to analyze the global asymptotic stability for the same power plants (Larderello, Toshiba and West Thurrock). The global asymptotic stability can be achieved for the power plants working in both the original and the restructured configurations. For example in Fig. \ref{fig:Toshlyap}, Toshiba power plant in its original configuration (red point) requires a higher temperature gradient to promote a heat flow capable of producing the working fluid works in cycles, in contrast to its operation in its restructured configuration (blue point). In fact, the speed of convergence to the respective steady states can be analyzed through $\dot{V}(x,y)$ for $k_{RC}$ and $k_{OC}$ as follows \cite{Jeffreys62}:

\begin{equation}
    M_V=\begin{array}{c}\lim\\\left(x,y\right)\rightarrow\left(\bar{x},\bar{y}\right)\end{array}\left[\frac{\dot{V}_{k_{OC}}\left(x,y\right)}{\dot{V}_{k_{RC}}\left(x,y\right)}\right]<1. \label{eq:limlypmT}
\end{equation}

\begin{figure}
    \centering
    \includegraphics[scale=0.9]{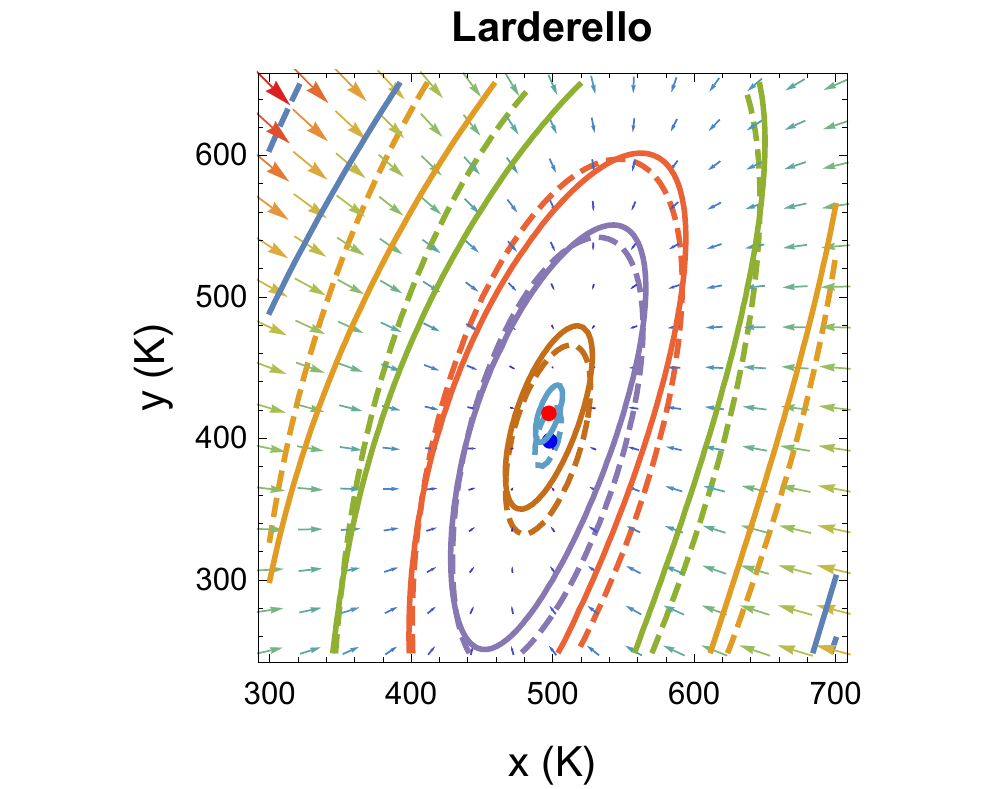}\includegraphics[scale=0.9]{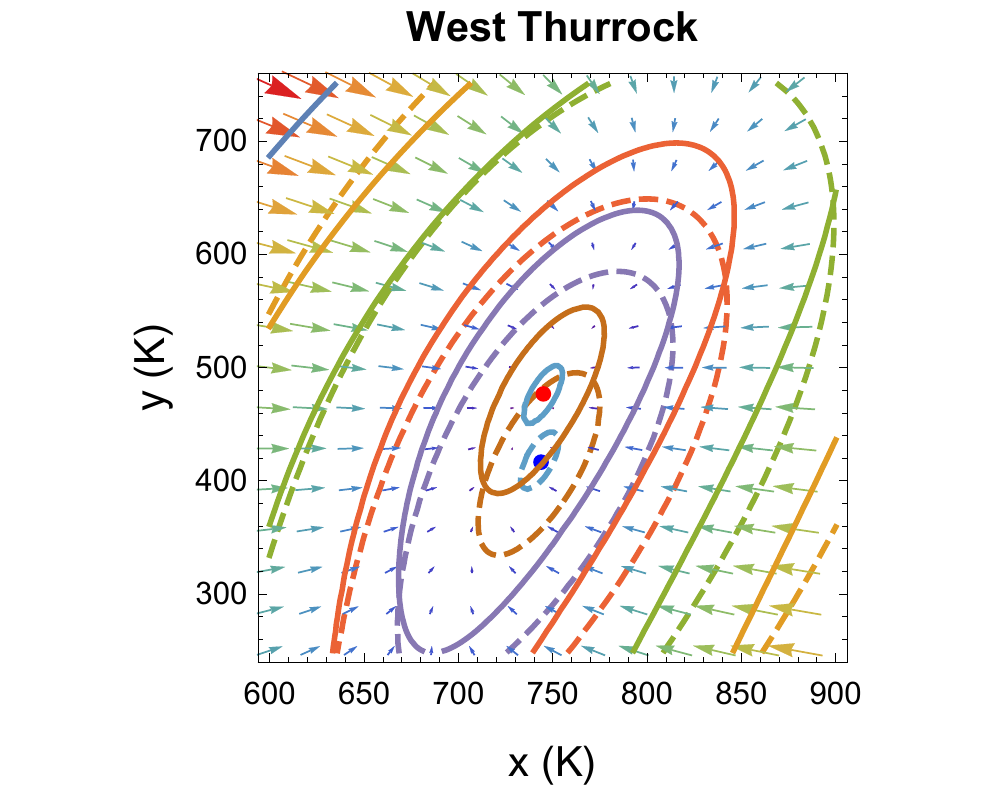}
    \caption{Qualitative plots of level curves of the Lyapunov function for the original (solid curves) and restructured (dashed curves) configurations for two real simple--cycle power plants (see Table \ref{tab:heatdata}). It is also see how their associated vector fields show that the steady states (blue and red points) are global asymptotically stables.}
    \label{fig:heng40}
\end{figure}

Analogously, in Fig. \ref{fig:heng40} we also show the qualitative level curves for Larderello and West Thurrock plants, their Lyapunov functions have a similar shape as Toshiba plant. Although both of power plants are of monocycle type $M_V <1$, this means that for any trajectory lying on a surface that represents the restructured configuration, it converges faster than the one of the original configuration.

Finally, we can observe in the three cases here analyzed, the level curves defined by $V(x,y)=\Omega$ for different values of the constant $\Omega$, show that as the constant value decreases, their level curves also decrease towards their corresponding steady-state values $(\bar{x},\bar{y})$.  

\section{Concluding remarks}
Every energy converter (heat engines or refrigerators mainly), under certain design and operating conditions, either operate on a very dissipative regime or on a reasonable efficient one. These energetic characteristics also affect the dynamic stability of the converters and can be reflected through a perturbative thermal analysis close to or far from the respective steady states. Although real heat engines are complex devices, the performance of their realistic upper bounds can be studied via relatively simple thermodynamic models, as is the case of FTT--models. This fact has been emphasized by other authors through very illustrative cases using simple FTT--models to describe some global properties of the energy converters. In the case of an endoreversible heat engine model and under the maximum $k$--efficient power regime, we found that operation modes with $-1<k<0$ decay asymptotically faster to the steady state than the ones with $k\geq0$. Likewise, we have found a Lyapunov function that corroborates the global asymptotic behavior of the steady states when the systems are linear. From Fig. \ref{fig:LyF1} we can guarantee the surfaces of Lyapunov have zero curvature when $k\rightarrow \infty$.

In this work, we applied the Lyapunov method (Krasovskii’s theorem) to characterize the global properties of a CA--heat engine model in terms of a Lyapunov function, which guarantees the global stability for certain parameter values related to control variables within the energy conversion scheme (see Fig. \ref{fig:CAeng40}). Moreover, our analysis of local and global asymptotically stability for the steady states was applied to three operating power plants, one of them belongs to the combined cycle type, the other two are of the simple--cycle type. In all the cases here studied, the internal temperatures approximate the steady--state values for the original and restructured configurations.

In the case of the local stability study, we observed that the simple--cycle power plants are less stable in the restructured configuration than in the original one, whereas for the combined cycle power plant happens the opposite. This behavior could be understood as follows: the more mechanical couplings a heat engine has to transform one type of energy into another, the better is its operation in a reconfigured energetic state. In the case of global stability analysis, we note that in general, the restructuring configurations are more stable than the original ones, the global stability is lost when the heat engines operate in low power conditions and whose efficiency is close to the reversible regime.

\end{document}